\pdfoutput=1

\documentclass[12pt,a4paper]{article}

\usepackage{ifthen} 
\newboolean{pdflatex}
\setboolean{pdflatex}{true} 

\newboolean{articletitles}
\setboolean{articletitles}{true} 

\newboolean{uprightparticles}
\setboolean{uprightparticles}{false} 

\def\paperauthors{LHCb collaboration} 
\def\paperasciititle{Observation of Xib0 to Xi0 Jpsi and evidence for Xib0 to Xi0 psi2S} 
\def\papertitle{Observation of $\Xibz \to \Xiz \jpsi$ \\ and evidence for $\Xibz \to \Xiz \psitwos$ decays} 
\def\paperkeywords{{High Energy Physics}, {LHCb}} 

\def\papercopyright{\the\year\ CERN for the benefit of the LHCb collaboration} 
\def\paperlicence{CC BY 4.0 licence}
\def\paperlicenceurl{https://creativecommons.org/licenses/by/4.0/}

\newif\ifEnableSectionTOCLinks
\EnableSectionTOCLinksfalse 

\usepackage[top=1in, bottom=1.25in, left=1in, right=1in]{geometry}

\usepackage{microtype}
\usepackage{lineno}  
\usepackage{xspace} 
\usepackage{caption} 

\usepackage{graphicx}  
\usepackage{color}
\usepackage{colortbl}
\graphicspath{{./figs/}} 

\usepackage{amsmath} 
\usepackage{amssymb}
\usepackage{amsfonts}
\usepackage{upgreek} 

\usepackage[normalem]{ulem} 

\newcommand*\patchAmsMathEnvironmentForLineno[1]{%
\expandafter\let\csname old#1\expandafter\endcsname\csname #1\endcsname
\expandafter\let\csname oldend#1\expandafter\endcsname\csname
end#1\endcsname
 \renewenvironment{#1}%
   {\linenomath\csname old#1\endcsname}%
   {\csname oldend#1\endcsname\endlinenomath}%
}
\newcommand*\patchBothAmsMathEnvironmentsForLineno[1]{%
  \patchAmsMathEnvironmentForLineno{#1}%
  \patchAmsMathEnvironmentForLineno{#1*}%
}
\AtBeginDocument{%
\patchBothAmsMathEnvironmentsForLineno{equation}%
\patchBothAmsMathEnvironmentsForLineno{align}%
\patchBothAmsMathEnvironmentsForLineno{flalign}%
\patchBothAmsMathEnvironmentsForLineno{alignat}%
\patchBothAmsMathEnvironmentsForLineno{gather}%
\patchBothAmsMathEnvironmentsForLineno{multline}%
\patchBothAmsMathEnvironmentsForLineno{eqnarray}%
}

\usepackage[pdftex,
            pdfauthor={\paperauthors},
            pdftitle={\paperasciititle},
            pdfkeywords={\paperkeywords}]{hyperref}
\usepackage{hyperxmp}
\hypersetup{
    pdfcopyright={Copyright (C) \papercopyright},
    pdflicenseurl={\paperlicenceurl}
}

\usepackage[colorinlistoftodos,textsize=scriptsize]{todonotes}

\usepackage[bottom,flushmargin,hang,multiple]{footmisc}

\usepackage[all]{hypcap} 

\usepackage{xspace} 
\usepackage{upgreek}

\def\lhcb   {\mbox{LHCb}\xspace}

\def\MagUp {\mbox{\em Mag\kern -0.05em Up}\xspace}

\ifdefined\ifuprightparticles
\else
\newboolean{uprightparticles}
\setboolean{uprightparticles}{false} 
\fi

\ifthenelse{\boolean{uprightparticles}}%
{
 
 \def\Pgamma      {\ensuremath{\upgamma}\xspace}

 \def\Pmu         {\ensuremath{\upmu}\xspace}

 \def\Ppi         {\ensuremath{\uppi}\xspace}

 \def\Ppsi        {\ensuremath{\uppsi}\xspace}

 \def\PDelta      {\ensuremath{\Delta}\xspace}                 
 \def\PXi         {\ensuremath{\Xi}\xspace}                 
 \def\PLambda     {\ensuremath{\Lambda}\xspace}                 
 \def\PSigma      {\ensuremath{\Sigma}\xspace}                 
 \def\POmega      {\ensuremath{\Omega}\xspace}                 
 \def\PUpsilon    {\ensuremath{\Upsilon}\xspace}
 \let\oldPi\Pi
 \def\PPi         {\ensuremath{\oldPi}\xspace}

 \def\PB      {\ensuremath{\mathrm{B}}\xspace}                 
 \def\PD      {\ensuremath{\mathrm{D}}\xspace}                 
 \def\PJ      {\ensuremath{\mathrm{J}}\xspace}                 
 \def\PK      {\ensuremath{\mathrm{K}}\xspace}                 
 \def\Pb      {\ensuremath{\mathrm{b}}\xspace}                 
 \def\Pc      {\ensuremath{\mathrm{c}}\xspace}                 
                  
 \def\Pe      {\ensuremath{\mathrm{e}}\xspace}                 
 \def\Pp      {\ensuremath{\mathrm{p}}\xspace}                 

 \def\Ps      {\ensuremath{\mathrm{s}}\xspace}

 \def\thebaroffset{0.0em}
}
{
 
 \def\Pgamma      {\ensuremath{\gamma}\xspace}

 \def\Pmu         {\ensuremath{\mu}\xspace}

 \def\Ppi         {\ensuremath{\pi}\xspace}

 \def\Ppsi        {\ensuremath{\psi}\xspace}                 
                  
 \mathchardef\PDelta="7101
 \mathchardef\PXi="7104
 \mathchardef\PLambda="7103
 \mathchardef\PSigma="7106
 \mathchardef\POmega="710A
 \mathchardef\PUpsilon="7107
 \mathchardef\PPi="7105
 \def\PB      {\ensuremath{B}\xspace}                 
 \def\PD      {\ensuremath{D}\xspace}                 
 \def\PJ      {\ensuremath{J}\xspace}                 
 \def\PK      {\ensuremath{K}\xspace}                 
 \def\Pb      {\ensuremath{b}\xspace}                 
 \def\Pc      {\ensuremath{c}\xspace}                 
                  
 \def\Pe      {\ensuremath{e}\xspace}                 
 \def\Pp      {\ensuremath{p}\xspace}                 

 \def\Ps      {\ensuremath{s}\xspace}

 \def\thebaroffset{0.18em}
}
\newcommand{\offsetoverline}[2][\thebaroffset]{\kern #1\overline{\kern -#1 #2}}%

\makeatletter
\ifcase \@ptsize \relax
  \newcommand{\miniscule}{\@setfontsize\miniscule{4}{5}}
\or
  \newcommand{\miniscule}{\@setfontsize\miniscule{5}{6}}
\or
  \newcommand{\miniscule}{\@setfontsize\miniscule{5}{6}}
\fi
\makeatother

\DeclareRobustCommand{\optbar}[1]{\shortstack{{\miniscule (\rule[.5ex]{1.25em}{.18mm})}
  \\ [-.7ex] $#1$}}

\def\epem       {{\ensuremath{\Pe^+\Pe^-}}\xspace}

\def\mup        {{\ensuremath{\Pmu^+}}\xspace}
\def\mun        {{\ensuremath{\Pmu^-}}\xspace} 

\def\mumu       {{\ensuremath{\Pmu^+\Pmu^-}}\xspace}

\def\squark    {{\ensuremath{\Ps}}\xspace}

\def\cquark    {{\ensuremath{\Pc}}\xspace}

\def\bquark    {{\ensuremath{\Pb}}\xspace}

\def\pion   {{\ensuremath{\Ppi}}\xspace}
\def\piz    {{\ensuremath{\pion^0}}\xspace}

\def\pim    {{\ensuremath{\pion^-}}\xspace}

\def\KorKbar {\kern \thebaroffset\optbar{\kern -\thebaroffset \PK}{}\xspace}

\def\D       {{\ensuremath{\PD}}\xspace}

\def\DorDbar {\kern \thebaroffset\optbar{\kern -\thebaroffset \PD}\xspace}

\def\Dp      {{\ensuremath{\D^+}}\xspace}
\def\Dm      {{\ensuremath{\D^-}}\xspace}

\def\DpDm    {\ensuremath{\Dp {\kern -0.16em \Dm}}\xspace}

\def\B       {{\ensuremath{\PB}}\xspace}

\def\BorBbar {\kern \thebaroffset\optbar{\kern -\thebaroffset \PB}\xspace}

\def\Bd      {{\ensuremath{\B^0}}\xspace}

\def\BdorBdbar {\kern \thebaroffset\optbar{\kern -\thebaroffset \Bd}\xspace}

\def\Bs      {{\ensuremath{\B^0_\squark}}\xspace}

\def\BsorBsbar {\kern \thebaroffset\optbar{\kern -\thebaroffset \Bs}\xspace}

\def\jpsi     {{\ensuremath{{\PJ\mskip -3mu/\mskip -2mu\Ppsi}}}\xspace}
\def\psitwos  {{\ensuremath{\Ppsi{(2S)}}}\xspace}

\def\Y#1S{\ensuremath{\PUpsilon{(#1S)}}\xspace}

\def\proton      {{\ensuremath{\Pp}}\xspace}

\def\Lz          {{\ensuremath{\PLambda}}\xspace}

\def\LorLbar     {\kern \thebaroffset\optbar{\kern -\thebaroffset \PLambda}\xspace}

\def\Xires       {{\ensuremath{\PXi}}\xspace}
\def\Xiz         {{\ensuremath{\Xires^0}}\xspace}
\def\Xim         {{\ensuremath{\Xires^-}}\xspace}

\def\Omegares    {{\ensuremath{\POmega}}\xspace}

\def\Xicp        {{\ensuremath{\Xires^+_\cquark}}\xspace}

\def\Lb           {{\ensuremath{\Lz^0_\bquark}}\xspace}

\def\Xibz         {{\ensuremath{\Xires^0_\bquark}}\xspace}
\def\Xibm         {{\ensuremath{\Xires^-_\bquark}}\xspace}

\def\Omegab       {{\ensuremath{\Omegares^-_\bquark}}\xspace}

\def\to                 {\ensuremath{\rightarrow}\xspace}

\def\AT#1     {\ensuremath{A_{\mathrm{T}}^{#1}}\xspace}           

\def\C#1      {\ensuremath{\mathcal{C}_{#1}}\xspace}                       
\def\Cp#1     {\ensuremath{\mathcal{C}_{#1}^{'}}\xspace}                    
\def\Ceff#1   {\ensuremath{\mathcal{C}_{#1}^{\mathrm{(eff)}}}\xspace}        
\def\Cpeff#1  {\ensuremath{\mathcal{C}_{#1}^{'\mathrm{(eff)}}}\xspace}       
\def\Ope#1    {\ensuremath{\mathcal{O}_{#1}}\xspace}                       
\def\Opep#1   {\ensuremath{\mathcal{O}_{#1}^{'}}\xspace}                    

\newcommand{\aunit}[1]{\ensuremath{\text{\,#1}}}       

\newcommand{\tev}{\aunit{Te\kern -0.1em V}\xspace}
\newcommand{\gev}{\aunit{Ge\kern -0.1em V}\xspace}
\newcommand{\mev}{\aunit{Me\kern -0.1em V}\xspace}
\newcommand{\kev}{\aunit{ke\kern -0.1em V}\xspace}
\newcommand{\ev}{\aunit{e\kern -0.1em V}\xspace}
 
\newcommand{\mevc}{\ensuremath{\aunit{Me\kern -0.1em V\!/}c}\xspace}
\newcommand{\gevc}{\ensuremath{\aunit{Ge\kern -0.1em V\!/}c}\xspace}
\newcommand{\mevcc}{\ensuremath{\aunit{Me\kern -0.1em V\!/}c^2}\xspace}
\newcommand{\gevcc}{\ensuremath{\aunit{Ge\kern -0.1em V\!/}c^2}\xspace}

\def\m    {\aunit{m}\xspace}

\def\fb   {\ensuremath{\aunit{fb}}\xspace}
\def\invfb   {\ensuremath{\fb^{-1}}\xspace}

\def\ns   {\ensuremath{\aunit{ns}}\xspace}

\newcommand{\stat}{\aunit{(stat)}\xspace}
\newcommand{\syst}{\aunit{(syst)}\xspace}

\newcommand{\chisq}{\ensuremath{\chi^2}\xspace}

\def\gsim{{~\raise.15em\hbox{$>$}\kern-.85em
          \lower.35em\hbox{$\sim$}~}\xspace}
\def\lsim{{~\raise.15em\hbox{$<$}\kern-.85em
          \lower.35em\hbox{$\sim$}~}\xspace}

\def\sPlot{\mbox{\em sPlot}\xspace}

\def\pt         {\ensuremath{p_{\mathrm{T}}}\xspace}

\def\evtgen     {\mbox{\textsc{EvtGen}}\xspace}

\def\geant      {\mbox{\textsc{Geant4}}\xspace}

\def\photos     {\mbox{\textsc{Photos}}\xspace}

\def\pythia     {\mbox{\textsc{Pythia}}\xspace}

\def\roofit     {\mbox{\textsc{RooFit}}\xspace}

\def\tell1  {TELL1\xspace}
\def\ukl1   {UKL1\xspace}

\newcommand{\lhcborcid}[1]{\href{https://orcid.org/#1}{\hspace*{0.1em}\raisebox{-0.45ex}{\includegraphics[width=1em]{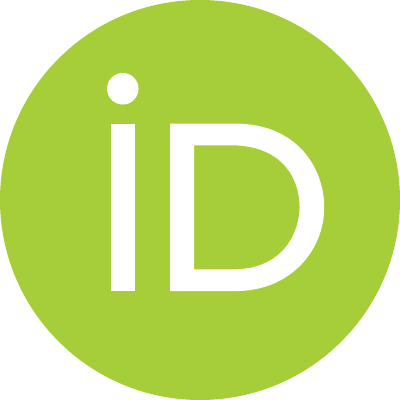}}}}

\hypersetup{
  colorlinks   = true, 
  urlcolor     = blue, 
  linkcolor    = blue, 
  citecolor    = red   
}

\ifEnableSectionTOCLinks
    \usepackage[explicit]{titlesec} 
    
    \let\oldcontentsline\contentsline
    \renewcommand

    \titleformat{\section}{\normalfont\Large\bf}{\hyperlink{tocsection.\thesection}{{\thesection} \parbox[t]{\dimexpr\textwidth-1pc}{#1}}}{1pc}{}

    \titleformat{\subsection}{\normalfont\bf}{\hyperlink{tocsubsection.\thesubsection}{{\thesubsection} \parbox[t]{\dimexpr\textwidth-1pc}{#1}}}{1pc}{}

    \titleformat{name=\section,numberless}[display]{}{}{0pt}{\normalfont\Huge\bfseries #1}
\fi

\usepackage{cite} 
\usepackage{LHCb/mciteplus}
\makeatletter
\g@addto@macro\bfseries{\boldmath}
\makeatother

\usepackage{longtable}

\begin{document}

\renewcommand{\thefootnote}{\fnsymbol{footnote}}
\setcounter{footnote}{1}

\begin{titlepage}
\pagenumbering{roman}

\vspace*{-1.5cm}
\centerline{\large EUROPEAN ORGANIZATION FOR NUCLEAR RESEARCH (CERN)}
\vspace*{1.5cm}
\noindent
\begin{tabular*}{\linewidth}{lc@{\extracolsep{\fill}}r@{\extracolsep{0pt}}}
\ifthenelse{\boolean{pdflatex}}
{\vspace*{-1.5cm}\mbox{\!\!\!\includegraphics[width=.14\textwidth]{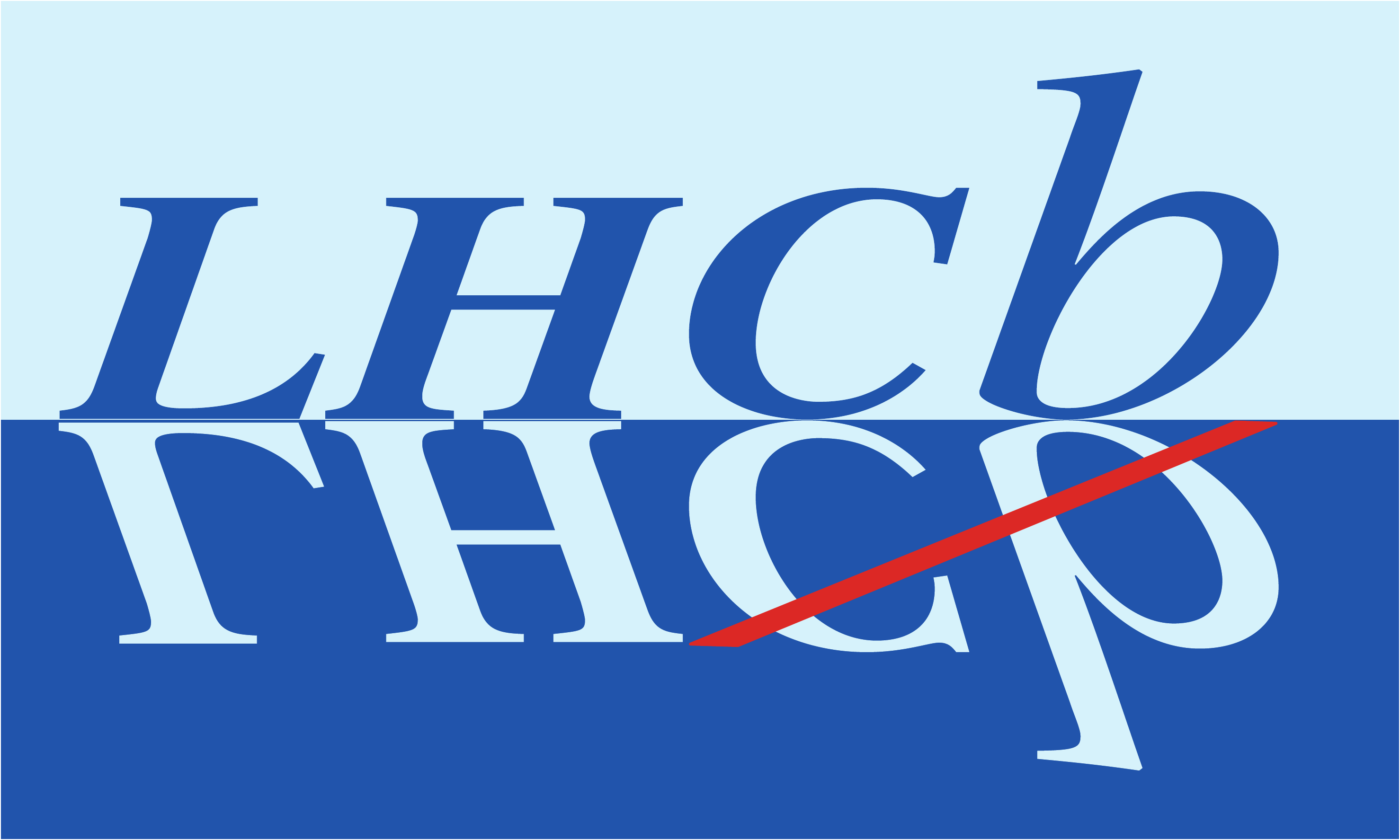}} & &}%
{\vspace*{-1.2cm}\mbox{\!\!\!\includegraphics[width=.12\textwidth]{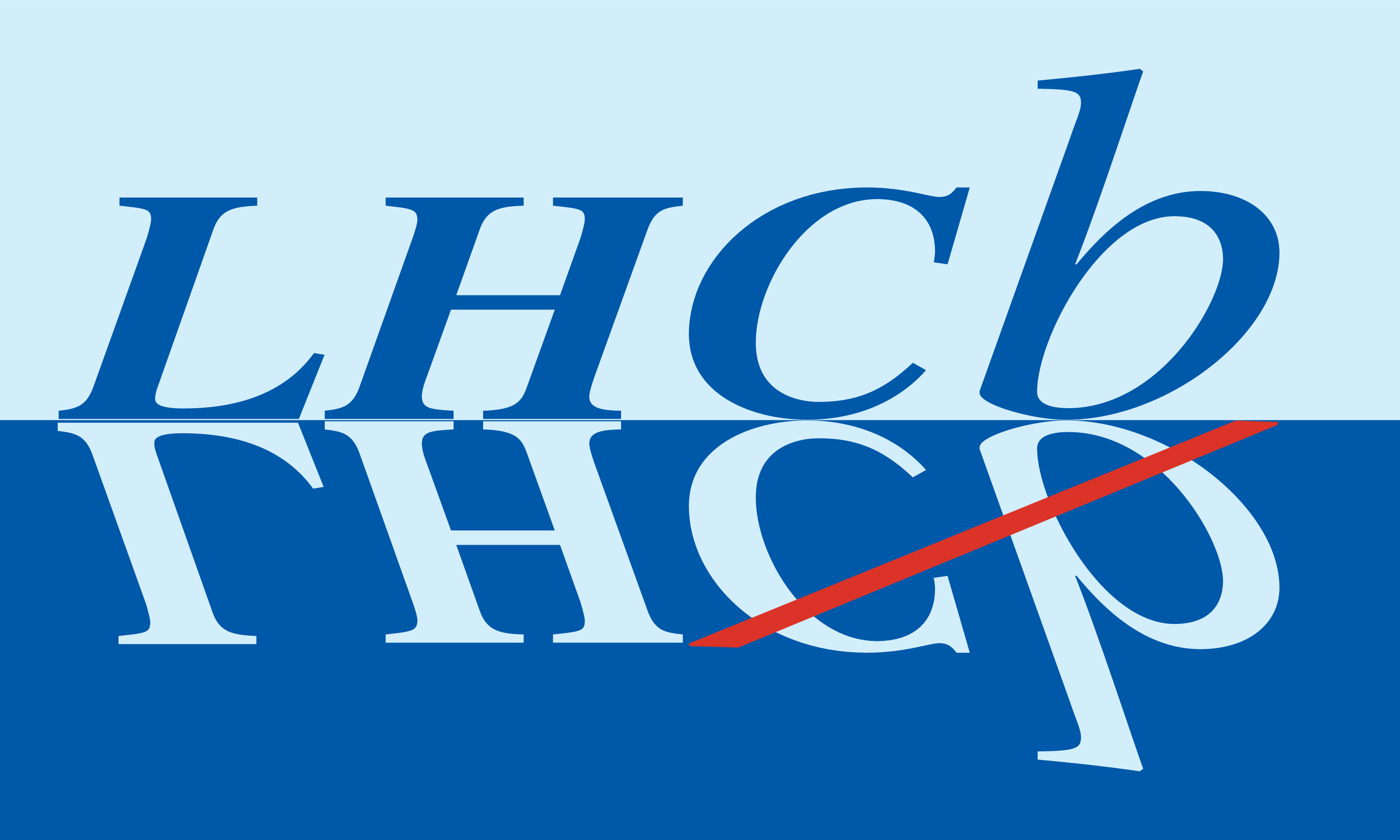}} & &}%
\\
 & & CERN-EP-2026-233 \\  
 & & LHCb-PAPER-2026-030 \\  
 & & 14 September 2026 \\
 & & \\
\end{tabular*}

\vspace*{4.0cm}

{\normalfont\bfseries\boldmath\huge
\begin{center}
  \papertitle 
\end{center}
}

\vspace*{2.0cm}

\begin{center}

\paperauthors\footnote{Authors are listed at the end of this Letter.}
\end{center}

\vspace{\fill}

\begin{abstract}
  \noindent
    The first search for the beauty baryon decays $\Xibz \to \Xiz \jpsi$ and $\Xibz \to \Xiz \psitwos$ is presented using the proton-proton collision dataset collected by the LHCb experiment between 2016 and 2018, corresponding to an integrated luminosity of $5.4\,\mathrm{fb}^{-1}$. The first observation of the decay $\Xibz \to \Xiz \jpsi$ is reported and evidence of the decay $\Xibz \to \Xiz \psitwos$ is presented. The \Xiz hyperon is fully reconstructed for the first time at an LHC experiment, which is achieved using the $\Xiz \to \Lz \piz$ decay. The ratio of the branching fractions is measured as
  \begin{equation*}
  \frac{\mathcal{B}(\Xibz \to \Xiz \psitwos)}{\mathcal{B}(\Xibz \to \Xiz \jpsi)} = 0.59 \pm 0.19 \stat \pm 0.04 \syst.
  \end{equation*} 
  
\end{abstract}

\vspace*{2.0cm}

\begin{center}
  Submitted to
  Phys.~Rev.~Lett.
\end{center}

\vspace{\fill}

{\footnotesize 
\centerline{\copyright~\papercopyright. \href{\paperlicenceurl}{\paperlicence}.}}
\vspace*{2mm}

\end{titlepage}


\newpage
\setcounter{page}{2}
\mbox{~}

\renewcommand{\thefootnote}{\arabic{footnote}}
\setcounter{footnote}{0}


\cleardoublepage


\pagestyle{plain} 
\setcounter{page}{1}
\pagenumbering{arabic}

Weakly decaying $b$ baryons offer a valuable testing ground for the Standard Model, providing a complementary perspective to the more extensively studied meson decays. There are four ground-state baryons containing exactly one $b$ quark but no other heavy quark and decay exclusively via the weak interaction: the $\Lb$ and $\Omegab$ baryons being isospin singlets, and the \Xibz and \Xibm baryons forming an isospin doublet. Among these, the $\Xibz \, (usb)$ baryon remains the least studied. It was first observed by the CDF collaboration in 2011~\cite{xib0_obs} using the decay $\Xibz \to \Xicp \pim$. Since then, more \Xibz decay channels have been observed by the LHCb collaboration~\cite{LHCb-PAPER-2013-056, LHCb-PAPER-2023-008, LHCb-PAPER-2023-017}.
Recent LHCb measurements yield a \Xibz mass of \mbox{$5791.12 \pm 0.60 \pm 0.45 \pm 0.24 \mevcc$}~\cite{LHCb-PAPER-2023-017}, with the uncertainties quoted throughout this Letter being statistical, systematic, and due to external inputs, respectively. The \Xibz production cross-section, a necessary quantity for an absolute branching fraction ($\mathcal{B}$) measurement, has not yet been directly measured.

This Letter presents the first search for the decays $\Xibz \to \Xiz \jpsi$ and $\Xibz \to \Xiz \psitwos$, using the proton-proton ($pp$) collision dataset collected by the LHCb experiment from 2016 to 2018 at a center-of-mass energy of $\sqrt{s} = 13 \tev$. Charge conjugation is implied throughout this Letter. The dataset corresponds to an integrated luminosity of $5.4 \invfb$ and is used to measure the ratio of branching fractions of the two decays, 
\begin{equation*}
R_{\Xibz} \equiv \frac{\mathcal{B}(\Xibz \to \Xiz \psitwos)}{\mathcal{B}(\Xibz \to \Xiz \jpsi)},
\end{equation*}
in which many systematic uncertainties, such as the one arising from the unknown \Xibz production fraction, cancel.

The signal processes are reconstructed via the decays $\psi \to \mumu$, where $\psi$ is either the \jpsi or \psitwos meson, and $\Xiz \to \Lz \piz$, with $\Lz \to \proton \pim$ and $\piz \to \Pgamma \Pgamma$. The charmonium decays into muon pairs provide clean experimental signatures. The reconstruction of the neutral \Xiz hyperon decay vertex is experimentally challenging, introducing a substantial source of background.

Both signal decays presented in this Letter have not yet been observed, with a previous LHCb measurement of the decay $\Xibz \to \Lz \jpsi$ modeling a joint \mbox{$\PXi_{b}^{0,-} \to \PXi^{0,-} \jpsi$} background component \cite{LHCb-PAPER-2019-039}. The decay $\Xibz \to \Xiz \jpsi$ is predicted to have a branching fraction of ${\mathcal{B}(\Xibz \to \Xiz \jpsi) = (4.4 \pm 0.9) \times 10^{-4}}$ in the covariant confined quark model~\cite{Gutsche:2018utw}. The perturbative QCD approach predicts a value of ${(7.5^{\:+\:1.2\, +0.3}_{\:-\:0.1\, -0.2}) \times 10^{-4}}$~\cite{Rui:2023fiz}. No theoretical predictions exist for the $\psitwos$ mode. The corresponding ratios for the \Lb baryon measured by the LHCb~\cite{LHCb-PAPER-2017-041} and ATLAS collaborations~\cite{lambdabtension_3}, ${R_{\Lb}= 0.513 \pm 0.023  \pm 0.016 }$ and ${R_{\Lb}= 0.501 \pm 0.033  \pm 0.019 }$, respectively, are in agreement with the value predicted by perturbative QCD, $R_{\Lb}= 0.47^{\:+\:0.02}_{\:-\:0.00}$~\cite{Rui:2022sdc}. Yet, the measurements disagree with the covariant quark model prediction of $R_{\Lb}= 0.8\pm 0.1$ at $2.8$ standard deviations ($\sigma$)~\cite{lambdabtension_2}. The CMS experiment has recently measured the corresponding ratio for the \Xibm baryon as $0.84^{\:+\:0.21}_{\:-\:0.19}  \pm 0.10 \pm 0.02$~\cite{CMS:2024rbi}. Under isospin symmetry, $R_{\Xibm}$ is expected to be equal to $R_{\Xibz}$.

The \lhcb detector~\cite{LHCb-DP-2008-001,LHCb-DP-2014-002} is a single-arm forward
spectrometer covering the \mbox{pseudorapidity} range $2<\eta <5$. The detector includes a high-precision tracking system, including a silicon-strip vertex detector surrounding the $pp$ interaction region. Muons are identified by a
system composed of alternating layers of iron and multiwire
proportional chambers~\cite{LHCb-DP-2012-002}. Photons, electrons, and neutral hadrons are identified by a calorimeter system.
The online event selection is performed by a trigger~\cite{LHCb-DP-2012-004}, which consists of a hardware and a software stage, the latter applying a full event reconstruction. Triggered data subsequently undergo a centralized offline processing step~\cite{Stripping}.

The decay vertex of the \Xibz baryon is reconstructed using the prompt decay of the $\psi$ states into a muon pair. The $\Xiz \to \Lz \piz$ decay vertex cannot be reconstructed directly since the \piz mesons do not have a defined trajectory. While the $\Lz \to \proton \pim$ decay is fully reconstructed, the \piz momentum vector is determined assuming the meson originates from the primary $pp$ collision vertex (PV) and does not take into account its displacement due to the long-lived \Xiz baryon, biasing its momentum measurement. Owing to this, the \Xiz momentum is determined with limited precision. The signal decay kinematics result in soft \piz mesons, leading to a large background. These effects on the \Xiz hyperon reconstruction are mitigated using a kinematic decay chain fit~\cite{Hulsbergen:2005pu} that constrains the \Xibz baryon to originate from the PV and the $\psi$, \Xiz, \Lz, and \piz masses to their known values~\cite{PDG2024}. The refitted four-momenta are used to determine the proper decay time of the \Xiz hyperon and the invariant mass $m(\Xiz \psi)$ of the \Xibz candidate, improving its mass resolution by a factor of two.

Simulation is used to model selection requirements and the associated efficiencies, as well as the expected background distributions. In the simulation, $pp$ collisions are generated using \pythia ~\cite{Sjostrand:2007gs} with a specific \lhcb configuration~\cite{LHCb-PROC-2010-056}. Decays of unstable particles are described by \evtgen~\cite{Lange:2001uf}, in which final-state radiation is generated using \photos~\cite{davidson2015photos}. The interaction of the generated particles with the detector, and its response, are implemented using the \geant toolkit~\cite{Allison:2006ve, *Agostinelli:2002hh} as described in Ref.~\cite{LHCb-PROC-2011-006}. The simulated samples are corrected for data-simulation differences in the \Xibz production kinematics using the more statistically abundant dataset of $\Xibm \to \Xim \jpsi$ decays, with $\Xim \to \Lz \pim$. A gradient-boosted weighting algorithm~\cite{Rogozhnikov:2016bdp} computes weights between simulation and data, where the latter is background-subtracted using the \sPlot method~\cite{Pivk:2004ty}. The muon identification and hardware trigger response are corrected using a tag-and-probe approach on dedicated calibration data samples~\cite{LHCb-PUB-2014-039, LHCb-DP-2018-001}.
After the calibration, good agreement between the $\Xibz \to \Xiz \jpsi$ simulation and background-subtracted data is observed.

At the hardware trigger stage, $\Xibz \to \Xiz \psi$ candidates are required to contain a muon with a high transverse momentum with respect to the beam axis (\pt) or a $\mumu$ pair with a high product of \pt. At the software trigger stage, the $\mumu$ pairs are required to have a mass consistent with the world-average $\psi$ masses~\cite{PDG2024} and significant displacement from the PV. The $\Xibz \to \Xiz \psi$ signal candidates are reconstructed by combining a \Xiz and a $\psi$ candidate and are required to be within $1600 \mevcc$ of the known \Xibz baryon mass~\cite{PDG2024}. The \Xibz decay vertex is required to be of good quality and compatible with originating from its associated PV, which is the PV that fits best to the flight direction of the \Xibz candidate. The $\psi$ candidates are reconstructed from two oppositely charged particles that are identified as muons. The $\mumu$ mass is required to be within $100 \mevcc$ of the known $\psi$ mass values~\cite{PDG2024}. In the offline selection, the \Xibz decay vertex is reconstructed using the $\psi$ candidates and must be significantly displaced from the associated PV. The \Xiz hyperon candidate is combined from a $\Lz$ and a \piz candidate and required to have a mass smaller than $1450 \mevcc$ and  $\pt > 200 \mevc$. The \chisq associated with the distance of closest approach between the \Xiz and $\Lz$ candidates is required to be small.
The $\Lz$ candidate is built from two oppositely charged particles assumed to be a proton and a pion that leave tracks in the tracking station, but not in the vertex detector, and must have a mass within $8 \mevcc$ of the known value~\cite{PDG2024}. This unique decay topology is created by the successive decays of two long-lived hyperons. The tracks forming the $\Lz$ candidate are required to originate from the same vertex which must be less than $2.3 \m$ away from the PV along the beam axis to ensure that the reconstruction is not affected by material interactions in front of the magnet. The cosine of the angle between the momentum vector of the $\Lz$ candidate and the vector pointing from the PV to the \Lz decay vertex is required to be positive, and the apparent \Lz decay time when its origin vertex is extrapolated back to the associated PV must be within 0 and 2 \ns.
The \piz candidates are required to have a mass lower than $250 \mevcc$ and $\pt>100 \mevc$. Photon candidates must meet photon identification criteria and have a $\pt > 70 \mevc$, which selects the \piz mesons and photons with the lowest \pt used in an LHCb measurement so far. Only photon pairs that form separate clusters in the electromagnetic calorimeter are considered, as this is a feature of low-momentum \piz mesons.

The topology of two successive long-lived hyperon decays is not imitable by $B$ meson decays, therefore no mesonic background is expected. Yet, several baryonic decays must be considered. The decay $\Lb \to \Lz \psi$ combined with a random \piz meson can pollute the signal region. A mass veto is imposed by rejecting all candidates for which the reconstructed $\Lz \psi$ mass falls within $25 \mevcc$ of the known $\Lb$ mass~\cite{PDG2024}. Two other sources of peaking background are identified and indistinguishable: they originate from misreconstructed signal decays or their charged \Xibm counterparts $\Xibm \to \Xim \psi$. In both cases, the pion originating from the $\PXi^{0,-}$ decay is not reconstructed and a random \piz meson or a random diphoton combination is added. A requirement is imposed on the product of the two photons' \pt to suppress the misreconstructed $\PXi_b^{0,-}$ background. 

The dominant source of background is of combinatorial nature, especially due to the challenging reconstruction of the $\piz \to \Pgamma \Pgamma$ decay in a high-multiplicity environment. A boosted decision tree~(BDT) classifier~\cite{Breiman,AdaBoost}, implemented in the TMVA toolkit~\cite{Hocker:2007ht,*TMVA4}, is used to suppress this source of background. The classifier is trained using corrected \Xibz \to \Xiz \jpsi simulation as a signal proxy and data in the \Xiz \jpsi mass sidebands, namely $4900 < m(\Xiz \jpsi)< 5400\mevcc$ and $6100 < m(\Xiz \jpsi) < 7500\mevcc$, as the background proxy. The classifier uses information about decay kinematics, topology, and vertex-fit quality. Among the most discriminating variables are the reconstructed masses of the \Xiz and \piz candidates, as well as the \pt of the \Lz hyperon.
The BDT hyperparameters are optimized using a grid search to maximize the area under the Receiver Operating Characteristic (ROC) curve.
No effects of overtraining are observed when performing a ten-fold cross-validation. The working point of the classifier is optimized for the \jpsi and \psitwos modes individually to maximize the figure of merit defined in Ref.~\cite{Punzi:2003bu}.
After the selection is applied to the data, 14\% of events in the \jpsi mode and 26\% in the \psitwos mode contain more than one \Xibz candidate, where the latter number is higher due to the softer \piz candidates present in this mode. In case multiple candidates remain, one is selected randomly.

\begin{figure}[tb]
  \begin{center}
    \includegraphics[width=0.49\linewidth]{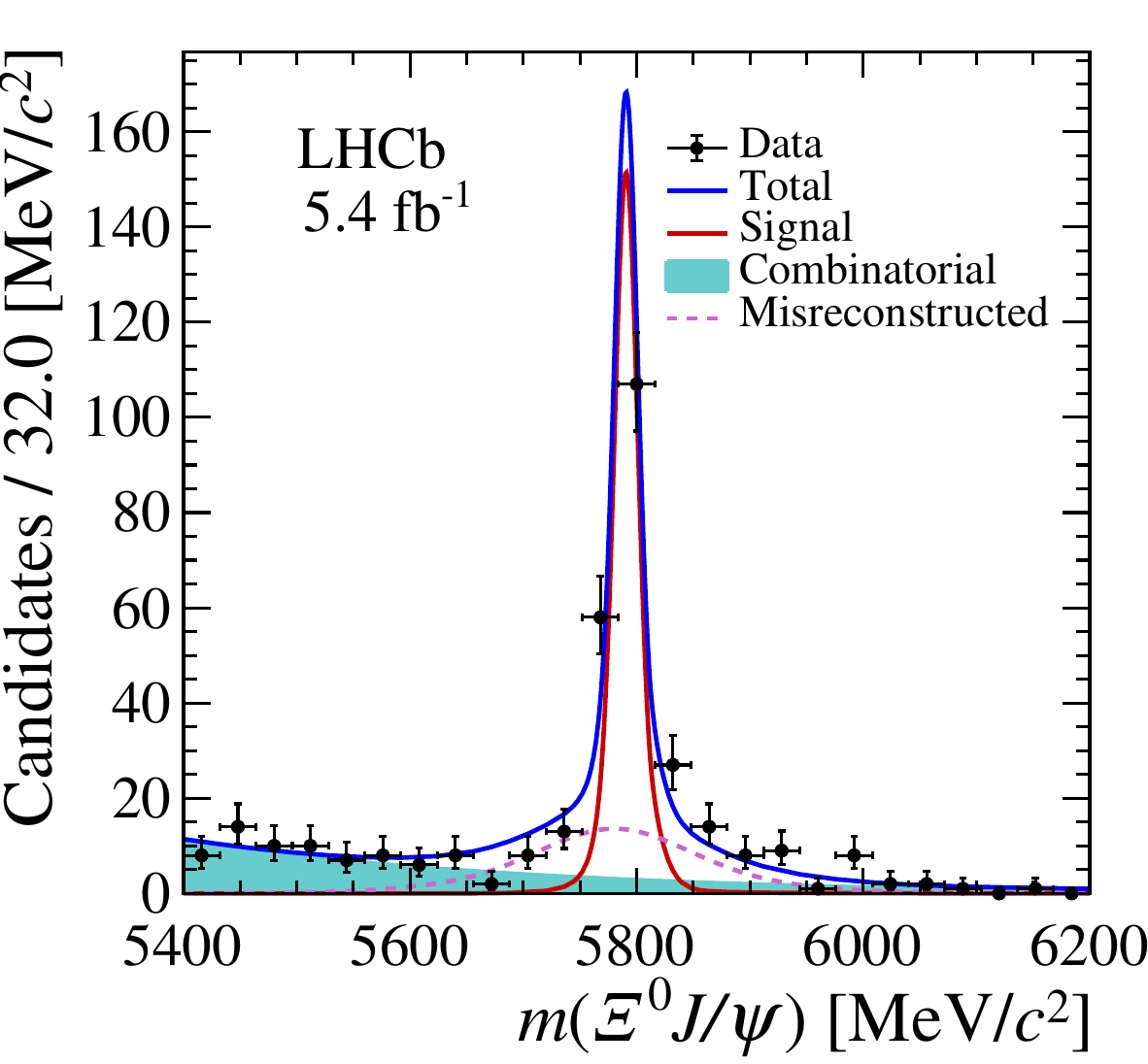}
    \includegraphics[width=0.49\linewidth]{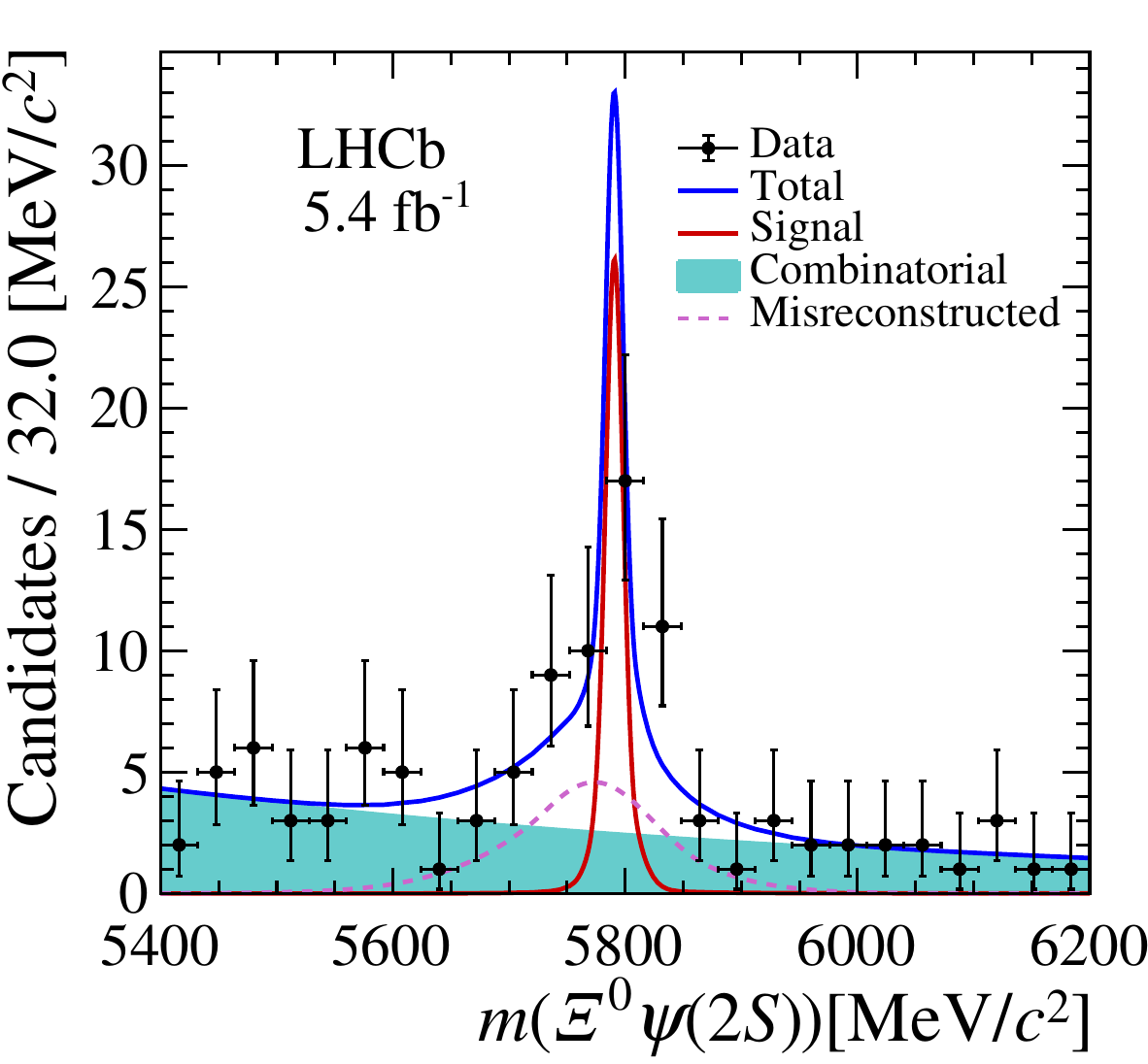}
    \vspace*{-0.5cm}
  \end{center}
  \caption{
    Invariant-mass distributions of (left) $\Xiz \jpsi$ and (right) $\Xiz \psitwos$ candidates, with the results of the simultaneous fit superimposed.}
  \label{fig:simfit}
\end{figure}

The yields of the $\Xibz \to \Xiz \jpsi$ and $\Xibz \to \Xiz \psitwos$ decays are determined using an extended unbinned maximum-likelihood fit to the reconstructed $\Xiz \psi$ invariant-mass distributions using the \roofit~\cite{roofit} package. The fit is performed in the signal mass window of $5400 < m(\Xiz \psi) < 6200 \mevcc$.  
Independent fits to each decay mode are performed to determine the statistical significances using likelihood scans. The \mbox{$\Xibz \to \Xiz \jpsi$} decay is observed for the first time, while evidence for the $\Xibz \to \Xiz \psitwos $ decay is seen with a significance exceeding $4 \sigma$. A simultaneous fit to the invariant masses $m(\Xiz \jpsi)$ and $m(\Xiz \psitwos)$ is performed to compute the branching-fraction ratio and is shown in Fig.~\ref{fig:simfit}. 

In the fits, the total probability density function is a linear combination of the individual fit components. Each signal is modeled using a linear combination of two one-sided Crystal Ball (CB) functions~\cite{dscb, Skwarnicki:1986xj} with a common mean, the tail parameters and combination fraction of which are fixed from simulation. The mean is shared between both signal modes in the simultaneous fit in order to improve the fit stability. Owing to the low expected yield in the \psitwos mode, the width of the signal peak cannot be reliably determined from data. A scale factor is determined from the measured mass resolution ratio between data and simulation for the \jpsi mode that is propagated to the \psitwos fit. Due to their similar kinematics, the misreconstructed $\PXi_b^{0,-}$ backgrounds cannot be distinguished in the $\Xibz$ mass spectrum and are merged in a common misreconstructed background to avoid overcounting the yield. This background is modeled using the misreconstructed signal simulation and fit with a linear combination of two double-sided CB functions. The parameters are fixed from simulation to improve the fit stability. The combinatorial background is modeled using an exponential function, the slope of which is allowed to vary freely.
All background yields, as well as the signal yield of the \jpsi component, are left free. The statistical uncertainty on $R_{\Xibz}$ is determined directly in the fit by letting this observable vary freely and expressing it using the signal yield of the \psitwos mode as
\begin{equation*}
N(\Xibz \to \Xiz \psitwos) = R_{\Xibz} \times N(\Xibz \to \Xiz \jpsi) \times \frac{\mathcal{B}(\psitwos \to \mup \mun)}{\mathcal{B}(\jpsi \to \mup \mun)} \times \frac{\epsilon_{\Xibz \to \Xiz \psitwos}}{\epsilon_{\Xibz \to \Xiz \jpsi}}.
\end{equation*}
The \jpsi signal yield is determined as $143 \pm 17$, from which the \psitwos signal yield is calculated as $18 \pm 7$. The branching-fraction ratio $\mathcal{B}(\psitwos \to \mup \mun)/\mathcal{B}(\jpsi \to \mup \mun)$ is taken from Ref.~\cite{PDG2024}. For the \psitwos decay mode, lepton universality is assumed and the more precise measurement of $\mathcal{B}(\psitwos \to \epem)$ is used. The efficiency ratio $\epsilon_{\Xibz \to \Xiz \psitwos}/\epsilon_{\Xibz \to \Xiz \jpsi}$ is computed using the corrected simulation and accounts for effects in the detector acceptance, decay reconstruction, and the candidate selection which is optimized for each decay mode. The efficiencies are computed as $ (9.30 \pm 0.10) \times 10^{-5}$ for the \jpsi mode and $(1.52 \pm 0.01) \times 10^{-4}$ for the \psitwos mode.

\begin{figure}[tb]
  \begin{center}
    \includegraphics[width=0.6\linewidth]{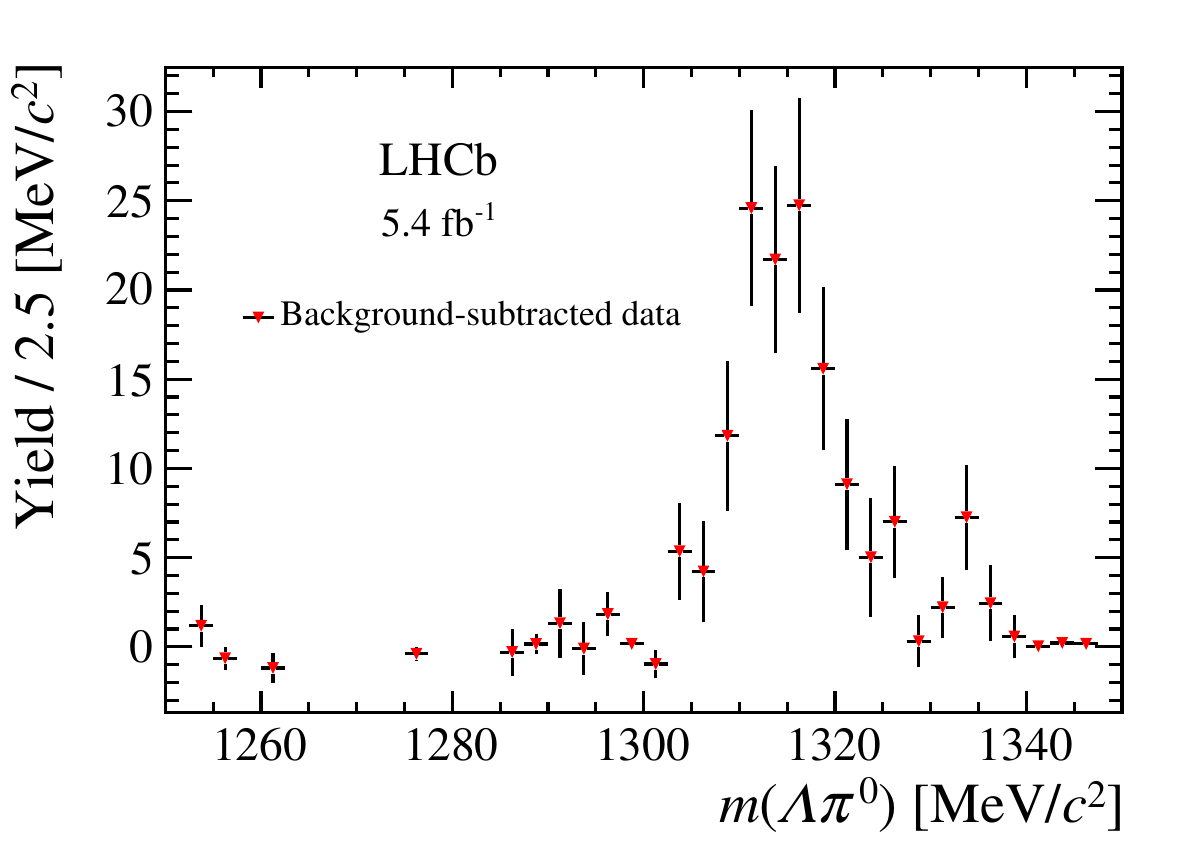}
  \end{center}
  \caption{
    Background-subtracted $\Lz \piz$ invariant-mass distribution in the \Xibz \to \Xiz \jpsi sample.}
  \label{fig:xi_mass}
\end{figure}

The observation of the $\Xibz \to \Xiz \jpsi$ mode marks the first full reconstruction of the \Xiz hyperon at an LHC experiment. The distribution of the \Xiz invariant mass calculated with the kinematic fit, $m(\Lz \piz)$, is shown in Fig.~\ref{fig:xi_mass} with the background  subtracted from the individual fit to the $\Xibz \to \Xiz \jpsi$ data using the \sPlot method. A check is performed using a multivariate selection that does not include the \Xiz and \piz invariant masses. A similar peak in the kinematically refitted \Xiz mass is seen, confirming the unbiased reconstruction of the hyperon.

\begin{table}[bt]
  \centering
  \caption{Relative systematic uncertainties on the observable $R_{\Xibz}$, in percent.} 
  \label{tab:total_syst_unc}
  \begin{tabular}{c c}
   \hline
    Source & Value\\
    \hline
    Simulation sample size & $1.3$\\
   Truth matching & $3.4$\\
    Calibration of simulation & $2.2$ \\
    Fit model & $4.2$ \\
    BDT selection & 3.0 \\
    Multiple candidate removal & 3.0 \\
    External branching fractions& $0.2$\\
    \hline
    Total & $7.3$\\
    \hline
 \end{tabular}
\end{table}

All systematic uncertainties considered in this Letter are listed in Table~\ref{tab:total_syst_unc}, with individual contributions summed in quadrature. The sample size of the simulated decays used to compute $R_{\Xibz}$ leads to a systematic uncertainty on the efficiency ratio. In the simulation, correctly reconstructed signal candidates are truth-matched to their counterparts in the generated decay chain. The change of $R_{\Xibz}$ with an alternative truth-matching procedure is taken as systematic uncertainty.

Several effects of the simulation correction are considered. The choice of hardware trigger selection used in the tag-and-probe method is changed, as well as the binning scheme of the corrections, giving a systematic uncertainty of $1.9\%$. The systematic uncertainty associated with the kinematic corrections arises from the limited calibration sample size and the weight-map binning. It is computed using the bootstrapping technique~\cite{efron:1979}, where each candidate in the calibration dataset is assigned a Poisson-distributed weight of unity mean. The root mean square (RMS) is $0.6\%$ which is assigned as systematic uncertainty.

The effect of the muon identification correction is assessed by varying the random-generator seed, kernel choices, and the track-multiplicity correction. For each setup, the total efficiency is recomputed, and the largest variation among these setups, $0.1\%$, is used to assign the systematic uncertainty.
No suitable calibration sample exists for the photons originating from the low-$\pt$ \piz candidates. The effect of the lack of correction is assessed by varying the requirement placed on the photon identification variable. The RMS is computed to be $1.0\%$ and assigned as systematic uncertainty.

Several effects of the fit model are studied using 20\,000 pseudoexperiments for each study. The fit bias is found to be $0.7\%$ and driven by the small measured signal yields. The impact of fixing the signal model tail parameters is evaluated by varying each by one standard deviation, yielding a maximum bias of 2.5\% which is assigned as the uncertainty.
The modeling of the misreconstructed background is studied using pseudo-experiments generated with misreconstructed \Xibm simulation samples which are fitted with the default model, yielding an uncertainty of $3.1\%$. Varying the mean and width of the misreconstructed background by one standard deviation each gives a systematic uncertainty of $1.3\%$.
The effect of the multivariate-selection optimization is studied by varying the BDT requirements around their default values. The resulting RMS is propagated as systematic uncertainty. Two alternative multiple-candidate removal methods are tested, retaining those candidates with reconstructed masses closer to the known \Xiz or \piz masses~\cite{PDG2024}. As $R_{\Xibz}$ is measured from the kinematically refitted \Xibz invariant mass, these methods do not introduce a bias. The RMS with respect to the default result is used as systematic uncertainty.
The external branching fractions of the charmonium decays propagate a systematic uncertainty of $0.2\%$ to the final result.

In summary, this Letter reports the first observation of the decay $\Xibz \to \Xiz \jpsi$ with an overwhelming statistical significance, and evidence for the decay $\Xibz \to \Xiz \psitwos$ at more than $4 \sigma$. Additionally, this measurement marks the first observation of a decay including a fully reconstructed \Xiz hyperon at the LHC with the softest photon and \piz meson candidates used in an LHCb measurement so far.
The analyzed dataset was collected in proton-proton collisions at $\sqrt{s} = 13 \tev$ by the LHCb experiment and corresponds to an integrated luminosity of $5.4 \invfb$.
The branching-fraction ratio between the decays $\Xibz \to \Xiz \psitwos$ and $\Xibz \to \Xiz \jpsi$ is measured as
\begin{equation*}
    R_{\Xibz} = \frac{\mathcal{B}(\Xibz \to \Xiz \psitwos)}{\mathcal{B}(\Xibz \to \Xiz \jpsi)} = 0.59 \pm 0.19 \pm 0.04,
\end{equation*}
where the quoted systematic uncertainty includes the uncertainty on the branching fractions of the charmonium decays.
The result reported in this Letter is compatible with the ratio recently measured by the CMS experiment for the \Xibm baryon~\cite{CMS:2024rbi}.
This result is also compatible with the corresponding ratio measured for the $\Lb$ baryon~\cite{LHCb-PAPER-2017-041, lambdabtension_3}. The observation of the decay \Xibz \to \Xiz \jpsi demonstrates that baryonic decays with multiple neutral particles can be reconstructed with good sensitivity at LHCb and implies that the rare decay $\Xibz \to \Xiz \mumu$ is within reach at LHCb with the increased data samples.

\section*{Acknowledgements}
%
%
\noindent We express our gratitude to our colleagues in the CERN
accelerator departments for the excellent performance of the LHC. We
thank the technical and administrative staff at the LHCb
institutes.
We acknowledge support from CERN and from the national agencies:
ARC (Australia);
CAPES, CNPq, FAPERJ and FINEP (Brazil); 
MOST and NSFC (China); 
CNRS/IN2P3 and CEA (France);  
BMFTR, DFG and MPG (Germany);
NKFIH (Hungary);              
INFN (Italy); 
NWO (Netherlands); 
MNiSW and NCN (Poland); 
MEC/IFA (Romania); 
MICIU and AEI (Spain);
SNSF and SER (Switzerland); 
NASU (Ukraine); 
STFC (United Kingdom); 
DOE NP and NSF (USA).
We acknowledge the computing resources that are provided by ARDC (Australia), 
CBPF (Brazil),
CERN, 
IHEP and LZU (China),
IN2P3 (France), 
KIT and DESY (Germany), 
INFN (Italy), 
SURF (Netherlands),
Polish WLCG (Poland),
IFIN-HH (Romania), 
PIC (Spain), CSCS (Switzerland), 
GridPP (United Kingdom),
and NSF (USA).  
We are indebted to the communities behind the multiple open-source
software packages on which we depend.
Individual groups or members have received support from
RTP (Australia), 
FWO Odysseus grant G0ASD25N (Belgium), 
Key Research Program of Frontier Sciences of CAS, CAS PIFI, CAS CCEPP (China); 
Minciencias (Colombia);
EPLANET, Marie Sk\l{}odowska-Curie Actions, ERC and NextGenerationEU (European Union);
A*MIDEX, ANR, IPhU and Labex P2IO, and R\'{e}gion Auvergne-Rh\^{o}ne-Alpes (France);
Alexander-von-Humboldt Foundation (Germany);
ICSC (Italy); 
Severo Ochoa and Mar\'ia de Maeztu Units of Excellence, GVA, XuntaGal, GENCAT, InTalent-Inditex and Prog.~Atracci\'on Talento CM (Spain);
the Leverhulme Trust, the Royal Society and UKRI (United Kingdom).

\addcontentsline{toc}{section}{References}

\bibliographystyle{LHCb/LHCb}
\bibliography{main,LHCb/standard,LHCb/LHCb-PAPER,LHCb/LHCb-CONF,LHCb/LHCb-DP,LHCb/LHCb-TDR,LHCb/LHCb-PUB}

\newpage
\centerline
{\large\bf LHCb collaboration}
\begin
{flushleft}
\small
R.~Aaij$^{39}$\lhcborcid{0000-0003-0533-1952},
M.~Abdelfatah$^{71}$,
A.S.W.~Abdelmotteleb$^{59}$\lhcborcid{0000-0001-7905-0542},
C.~Abellan~Beteta$^{53}$\lhcborcid{0009-0009-0869-6798},
F.~Abudin\'en$^{61}$\lhcborcid{0000-0002-6737-3528},
T.~Ackernley$^{63}$\lhcborcid{0000-0002-5951-3498},
A.A.~Adefisoye$^{71}$\lhcborcid{0000-0003-2448-1550},
B.~Adeva$^{49}$\lhcborcid{0000-0001-9756-3712},
M.~Adinolfi$^{57}$\lhcborcid{0000-0002-1326-1264},
P.~Adlarson$^{87,44}$\lhcborcid{0000-0001-6280-3851}
% ...

{\footnotesize \it

% ...
$^{39}$Nikhef National Institute for Subatomic Physics, Amsterdam, Netherlands\\
% ...
$^{44}$National Center for Nuclear Research (NCBJ), Warsaw, Poland\\
% ...
$^{49}$Instituto Galego de F{\'\i}sica de Altas Enerx{\'\i}as (IGFAE), Universidade de Santiago de Compostela, Santiago de Compostela, Spain\\
% ...
$^{53}$Physik-Institut, Universit{\"a}t Z{\"u}rich, Z{\"u}rich, Switzerland\\
% ...
$^{57}$H.H. Wills Physics Laboratory, University of Bristol, Bristol, United Kingdom\\
% ...
$^{59}$Department of Physics, University of Warwick, Coventry, United Kingdom\\
% ...
$^{61}$School of Physics and Astronomy, University of Edinburgh, Edinburgh, United Kingdom\\
$^{62}$School of Physics and Astronomy, University of Glasgow, Glasgow, United Kingdom\\
$^{63}$Oliver Lodge Laboratory, University of Liverpool, Liverpool, United Kingdom\\
% ...
$^{71}$Syracuse University, Syracuse, NY, United States\\
% ...
$^{87}$Department of Physics and Astronomy, Uppsala University, Uppsala, Sweden, associated to $^{62}$\\
% ...
}
\end{flushleft}

\end{document}